\documentclass[
  aps,prb,preprint,
  superscriptaddress,
  floatfix,
  longbibliography
]{revtex4-2}

\usepackage[utf8]{inputenc}
\usepackage[T1]{fontenc}
\usepackage{amsmath,amssymb,amsfonts,bm,mathrsfs}
\usepackage{braket,dsfont,dcolumn,graphicx}
\usepackage{xcolor}
\usepackage{comment}

\DeclareMathOperator{\sech}{sech}
\DeclareMathOperator{\Tr}{Tr}

\allowdisplaybreaks

\usepackage[unicode=true,pdfencoding=auto,
  colorlinks=true,linkcolor=blue,citecolor=blue,urlcolor=blue]{hyperref}

\begin{document}

\author{L.~F.~Alves~da~Silva}
\affiliation{Instituto de Física de São Carlos, Universidade de São Paulo, Caixa Postal 369, 13560-970, São Carlos - SP, Brazil}

\author{H.~Sanchez}
\affiliation{Instituto de Física de São Carlos, Universidade de São Paulo, Caixa Postal 369, 13560-970, São Carlos - SP, Brazil}

\author{M.~A.~Ponte}
\affiliation{Instituto de Física de São Carlos, Universidade de São Paulo, Caixa Postal 369, 13560-970, São Carlos - SP, Brazil}

\author{M.~H.~Y.~Moussa}
\affiliation{Instituto de Física de São Carlos, Universidade de São Paulo, Caixa Postal 369, 13560-970, São Carlos - SP, Brazil}

\author{Norton~G.~de~Almeida}
\affiliation{Instituto de Física, Universidade Federal de Goiás, 74.001-970, Goiânia - GO, Brazil}
\affiliation{Instituto de Física de São Carlos, Universidade de São Paulo, Caixa Postal 369, 13560-970, São Carlos - SP, Brazil}

\title{Superradiance and Superabsorption Engine of \(N\) Two-Level Systems:\\ \(N^{2}\)-Power Scaling at Near-Unity Efficiency}


\begin{abstract}

We present a thermal engine that exploits the \emph{cooperative superradiance} and \emph{superabsorption} of a sample of \(N\) two-level atoms. This engine operates using a single cold reservoir via cycles of collective pumping followed by decay. Using an effective mean-field Hamiltonian to describe the many-body dynamics, we design optimized drive pulses that preserve adiabaticity and achieve an average power output scaling quadratically with the system size, \(P \propto N^2\). An experimentally measurable figure of merit demonstrates that the efficiency of this superengine can approach unity. The resulting analytical model, which yields a representative Hamiltonian for the sample within the mean-field formalism, is validated by numerical simulations. Our results pave the way for scalable and highly efficient quantum heat engines based on collective effects.
\end{abstract}

\maketitle
\section{Introduction}
Since Dicke first predicted in 1954 that an ensemble of $N$ identical two-level emitters could radiate cooperatively~\cite{dicke1954coherence}, with an intensity scaling as $N^{2}$ rather than $N$, \emph{superradiant} (SR) decay has been confirmed across a wide variety of physical platforms—including atomic vapors, Rydberg gases, solid-state spins, and circuit-QED resonators~\cite{skribanowitz1973observation, gross1976observation, scheibner2007superradiance, rohlsberger2010collective, devoe1996observation, mlynek2014observation, kim2022photonic}. A landmark realization of steady-state superradiant emission was achieved in 2012~\cite{bohnet2012steady} using a Raman superradiant laser operating in the bad-cavity regime with fewer than one intracavity photon. More recently, the role of entanglement and mutual information in superradiance has been explored by coupling ensembles of two-level emitters either to a squeezed reservoir or to a one-dimensional waveguide~\cite{zhang2025unraveling}.

The mirror-image process, \emph{superabsorption} \cite{higgins2014superabsorption, mirzaei2015superabsorption, brown2019light, raimond1982collective}, can be achieved by engineering the environment or introducing suitable interactions to obtain an increase with $N^{2}$ of energy uptaking \cite{burgess2025engineering}.  
The deliberate combination of pulsed superradiant emission and controlled superabsorption therefore opens a new route to cooperative quantum machines whose work output and power can grow quadratically with system size while retaining the thermodynamic advantages of a cyclic engine. For example, in Ref.\cite{hardal2015superradiant} the authors propose an Otto cycle in which atoms prepared in coherent superposition states pass through a cavity while superradiating into the trapped field, allowing the extraction of power that scale with $N^{2}$. A shortcoming of this proposal is that the many injections of atoms lengthen the thermalization stage, decreasing the power. Furthermore, the maximum efficiency is that of Otto, which is lower than that of Carnot.  Refs.\cite{kamimura2022quantum} and \cite{kloc2019collective}  take advantage of superabsorption and superradiance, respectively, and propose a quantum Otto engine with improved power. Collective effects make the power scale with $N^{2}$, although the efficiency remains the same as that of a conventional Otto. Superabsorption can also be used to improve the performance of refrigerators Ref.\cite{kloc2021superradiant}.

It should be noted that in conventional quantum cycles, such as quantum Otto, both superabsorption and superradiance are considered strokes where heat exchange occurs without work being performed. The work is then extracted by changing the frequency of the sample oscillators, resulting in a power gain scaling as $N^{2}$ while keeping the efficiency below that Carnot cycle. In this work, we propose a prototype of \emph{superengine} in which both the superradiance and the superabsorption are part of the unitary expansion and compression strokes, allowing the extracted power to scale with $N^{2}$, respectively, while keeping the efficiency close to unity. To this end, we unified the mean-field treatment for both the absorption \cite{lorenzen2009quantum} and emission processes \cite{mizrahi1993pulsed}, which allowed us to treat these two phenomena as part of a process fast enough such that heat exchanges are neglected.

\section{Physical set--up} 
Consider an ensemble of $N$ identical two-level quantum emitters. These emitters can be implemented in a general way in a number of platforms, but for our purpose we name a few, such as  solid-state platforms \cite{liu2024super}, superconducting qubits in a low-$Q$ microwave resonator \cite{lambert2016superradiance},
or semiconductor quantum dots embedded in photonic or plasmonic cavities \cite{tiranov2023collective,wei2021plasmon}. 
We model each emitter acting effectively as a spin–$\tfrac12$ system with dipole moment $\boldsymbol{\mu}=\gamma\hbar\boldsymbol{\sigma}/2$, where $\gamma$ is the gyromagnetic ratio and $\boldsymbol{\sigma}$ denotes the Pauli vector. 
In these realizations, the sample is moderately dense with effective dimension $L$, and is engineered to satisfy $L\ll\lambda_0$, with $\lambda_0=2\pi c/\omega_0$ the resonant wavelength, ensuring that all emitters experience a nearly identical electromagnetic phase \cite{gross1982superradiance}. 
The energy gap $\hbar\omega_0$ between the two levels can be tuned either by an external bias field $B_0\hat{\mathbf{z}}$ (in the case of spin or qubit systems) or by optical/electrical controls (in quantum dots and plasmonic resonators). 
Depending on the engineered reservoir and pumping conditions, the mean field Hamiltonian $H_{MF}(t)$ drives the system toward collectively enhanced emission (superradiance) or collectively enhanced absorption (superabsorption), with effective rates $\gamma_{\mathrm{down}}$ and $\gamma_{\mathrm{up}}$, respectively, as detailed in Appendix B:
\begin{equation}
H_{MF}(t) \;=\; \frac{\omega_0}{2}\,\sigma_{z}
\;+\;\frac{N}{2}\,(\gamma_{\rm down}-\gamma_{\rm up})
\bigl(\langle\sigma_{x}\rangle\,\sigma_{y}
- \langle\sigma_{y}\rangle\,\sigma_{x}\bigr),
\label{eq:hamiltonian}
\end{equation}

where in Bloch representation:

\,

$\langle\sigma_x\rangle = r\sin\theta\cos\phi,\quad \langle\sigma_y\rangle = r\sin\theta\sin\phi,\quad \langle\sigma_z\rangle = r\cos\theta$, or explicitly:
\begin{align}
\langle \sigma_x(t) \rangle
&= r \,\sech\!\Bigl(\tfrac{t - t_d}{\tau}\Bigr)
  \cos\bigl(\phi_0 + \omega_0 t\bigr),
  \label{eq:sigmax} \\[6pt]
\langle \sigma_y(t) \rangle
&= r \,\sech\!\Bigl(\tfrac{t - t_d}{\tau}\Bigr)
  \sin\bigl(\phi_0 + \omega_0 t\bigr),
  \label{eq:sigmay} \\[6pt]
\langle \sigma_z(t) \rangle
&= -r \,\tanh\!\Bigl(\tfrac{t - t_d}{\tau}\Bigr),
  \label{eq:sigmaz}
\end{align}
where $\phi_0$ is the initial phase, which we choose to be zero,  $\omega_0$ is the atomic transition frequency, $t_d = \tau \ln\left(\cot(\theta_0/2)\right)$ is the delay time, marking the peak of the pulse, $\tau =2/{rN|\gamma_{\rm downn} - \gamma_{\rm up}|}$, $\gamma_{\rm down} \neq \gamma_{\rm up}$, is the characteristic time scale for superradiance ($\gamma_{\rm down} > \gamma_{\rm up}$, $\tau = \tau_{\rm emis}$, $t_d=t_{d}^{\rm emis}$) or superabsorption ($\gamma_{\rm up} > \gamma_{\rm down}, \tau=\tau_{\rm abs}$, $t_d=t_{d}^{\rm abs}$). The scaling factor $r$ locates the eigenstates of $H_{MF}(t)$ on the Bloch sphere oriented by angles $(\theta_{0},\phi_{0})$ and accounts for possible mixed-state effects, typically $r = 1$ for pure states. This effective Hamiltonian describes both a superradiant and a superabsorption pulse, whose intensity scales with $N^2$ - see Appendix B

\begin{equation}
\mathcal{I}(t) = -N\frac{d\varepsilon(t)}{dt}
= \left(\frac{N}{2}\right)^2 r\bigl(\gamma_{\rm down} - \gamma_{\rm up}\bigr)\omega_{0}\,\sech^2\!\Bigl(\frac{t - t_{d}}{\tau}\Bigr).
\label{eq:intensidade_superabsorcao}
\end{equation}

This approach simplifies the collective superradiance problem, which in Dicke’s formulation requires a Hilbert space that grows exponentially with $N$, into a problem of a single atom coupled to a self-consistent mean field. Although it is an effective model, it accurately captures the key transient features of superradiance before atomic relaxation occurs: the characteristic delay time, the peak intensity, and the shape of a $\mathrm{sech}^2$ pulse. Consequently, it provides a useful framework for investigating both superradiant and superabsorption bursts. Importantly, as demonstrated in Appendix B, this pulse is achieved almost unitarily, that is, without significant heat exchange, and becomes increasingly accurate as the number of emitters $N$ grows.

\begin{figure}[htbp]
  \centering
\includegraphics[width=\columnwidth]{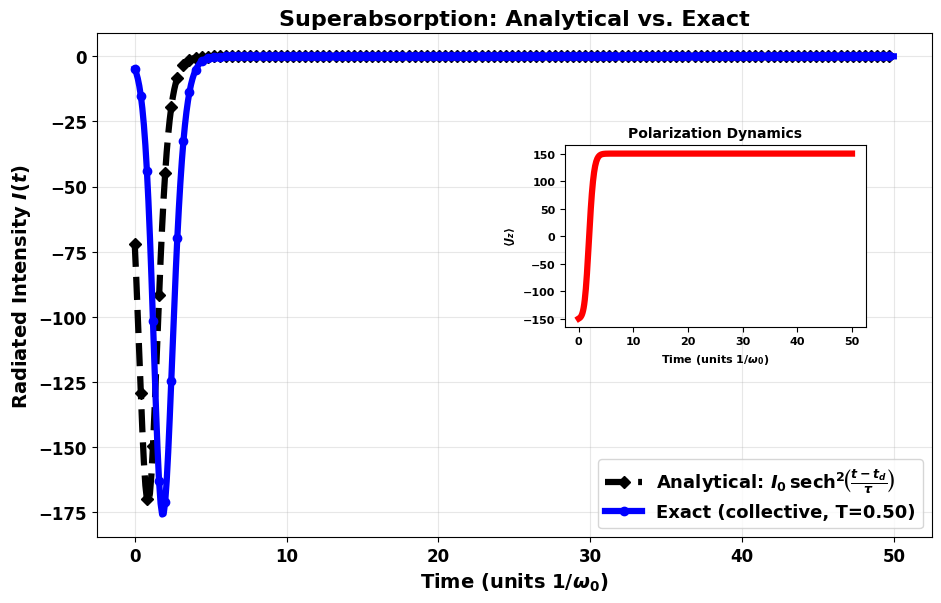}

  \caption{\text{Superabsorption}. Comparison of the analytical pulse 
\(I_0\,\sech^2\!\bigl(\tfrac{t - t_d}{\tau}\bigr)\) (dashed-black line) and the simulated (exact) absorbed intensity (solid-blue line), obtained from an initial thermal states undergoing collective decay. Input parameters: \(N=300\), \(\gamma_{\mathrm{up}}=0.01\), \(T=0.5\). Derived parameters:  \(t_d=28.9559\), $r = 0.7616$, $\theta_0 = 0.7050$ rad,  and \(\tau=0.8754\). The inset shows the polarization dynamics \(\langle J_z(t)\rangle\) versus. $t$. Time is given in units of $1/\omega_0$.}
\label{superabsortion}
\end{figure}

\begin{figure}[htbp]
  \centering
  \includegraphics[width=\columnwidth]{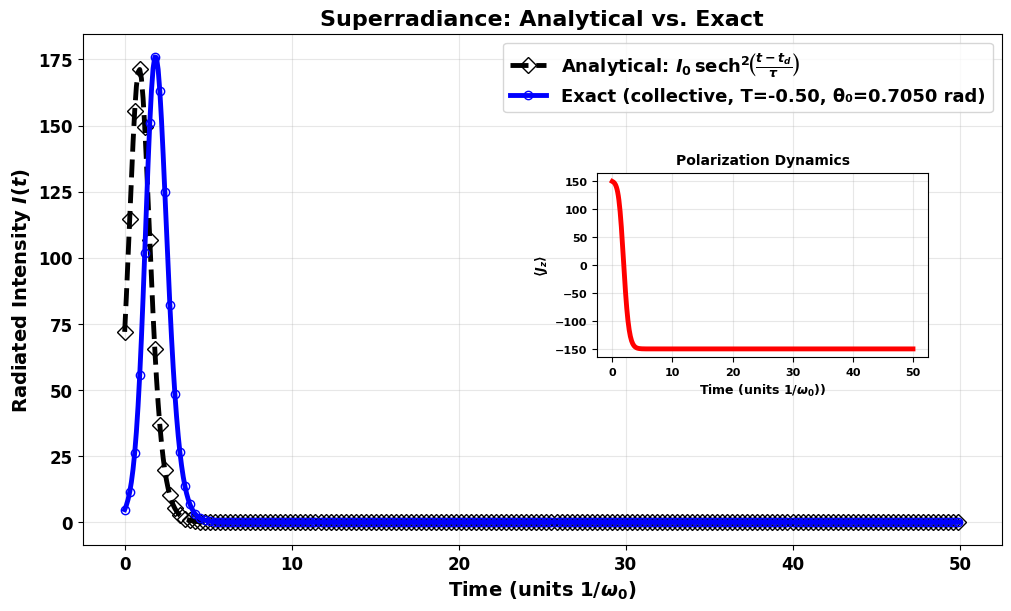}
 \caption{Superradiance. Comparison of the analytical pulse 
\(I_0\,\sech^2\!\bigl(\tfrac{t - t_d}{\tau}\bigr)\) (dashed black line with diamond markers) and the simulated radiated intensity (solid blue line with circle markers) starting from an initial thermal state undergoing collective absorption. Input parameters: \(N=300\), \(\omega_0=1.0\), \(\gamma_{\mathrm{down}}=0.01\), \(T = -0.5\). Since the population is inverted due to the collective pumping, we performed this simulation using a Gibbs state with effective negative temperature. Derived parameters: \(t_d = 0.8754\), \(r = 0.7616\), \(\theta_0 = 0.7050\) rad, and \(\tau = 0.8754\). The inset shows the polarization dynamics \(\langle J_z(t) \rangle\) versus \(t\). Time is given in units of \(1/\omega_0\).
}

\label{superradiance}

\end{figure}

To validate the analytical model against the exact dynamics, we performed numerical simulations using QuTiP \cite{johansson2012qutip} up to $N = 500$. The comparison is shown in Figs.~\ref{superabsortion} and~\ref{superradiance}, where the analytical (mean-field) result is shown by a dashed curve (black diamonds) and the exact numerical simulation is shown by a solid curve (blue circles). Using units of $\omega_0$, the parameters for superabsorption (Fig.~\ref{superabsortion}) are $\gamma_{\mathrm{up}}=0.01$ and $T=0.5$, resulting in $r = 0.7616$, $\theta_0 = 0.7050$ rad, $\tau = 0.8754$, and $t_d = 0.8754$. These exact (numerical) parameters are very close to those calculated using the mean-field formulas; see Appendix C. For superradiance (Fig.~\ref{superradiance}), the initial state must have an inverted population. We therefore choose the same decay rate $\gamma_{\mathrm{up}}=0.01$ and an effective negative temperature of $T = -0.5$. The other parameters, calculated from the same mean-field formulas, result in the values $r = 0.7616$, $\theta_0 = 0.07050$ rad, $\tau = 0.8754$, and $t_d = 0.8754$.

\begin{figure}[htbp]
  \centering
  \includegraphics[width=0.8\textwidth]{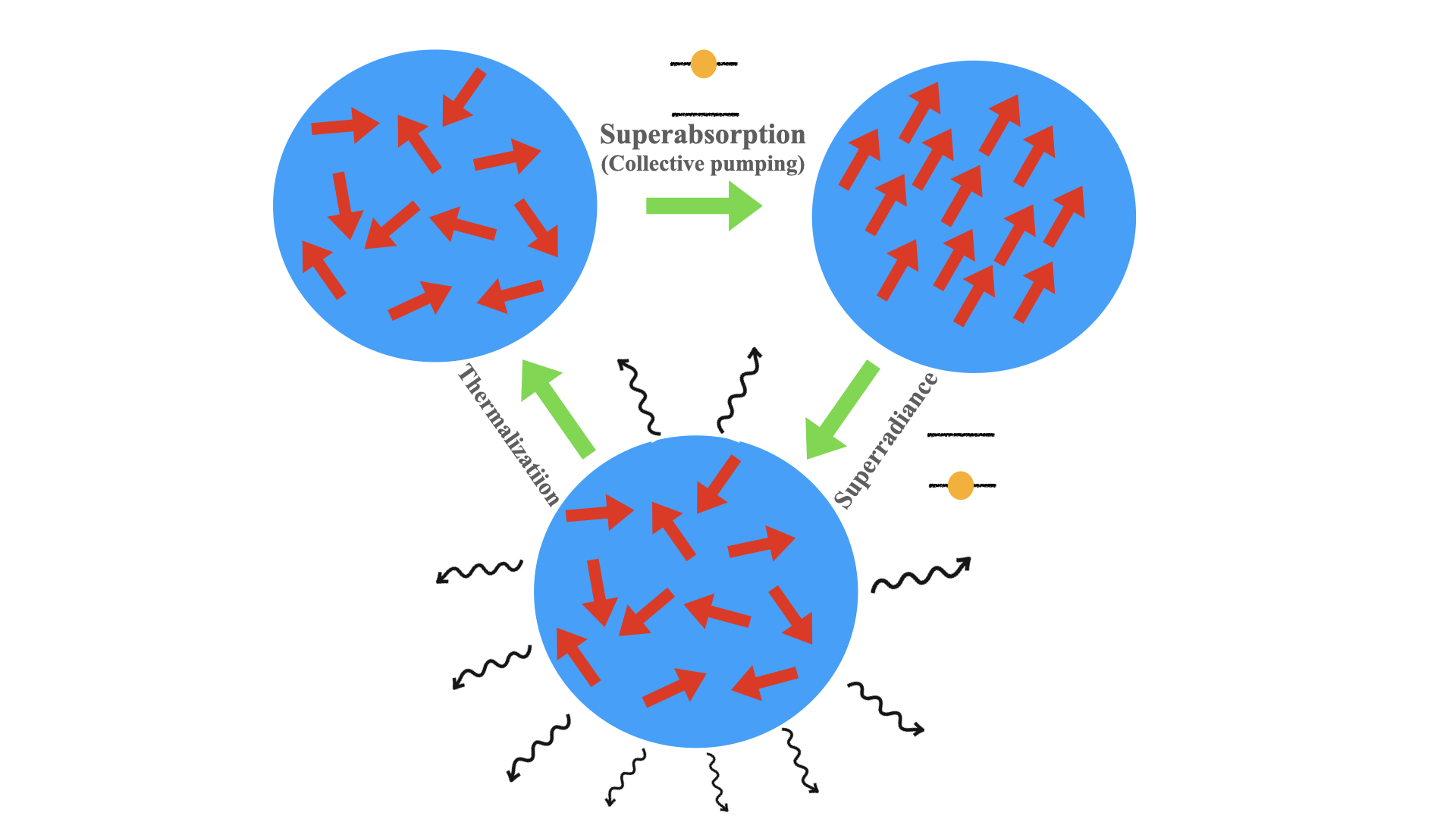}
  \caption{Superengine Cycle: The atomic sample begins in a thermal state with disordered dipole moments. It then undergoes multimodal pumping, triggering a superabsorption effect that inverts its population and orders the dipoles. This is followed by a superradiance emission, after which the sample re-thermalizes with the reservoir.}
  \label{scheme}
\end{figure}
\,

\section{Superengine Cycle.} The proposed quantum thermodynamic cycle consists of two unitary strokes (compression and expansion) and one isochoric stroke (thermalization). The time-dependent Hamiltonian for the unitary strokes is given by Eq.(\ref{eq:hamiltonian}). The cycle begins with the system thermalized in the ground state of the cold reservoir. For simplicity, we choose this initial state to be thermal. This initial thermalization serves as an ignition step and can be replaced by collectively exciting or de-exciting the sample, in which case the work done has to be account. Subsequently, a pump pulse is applied for a sufficiently short duration to achieve an inversion of the population with negligible heat exchange. Once the pump is turned off, the system undergoes superradiance, emitting its energy before returning to the thermal state with the reservoir, thus completing the cycle (see Fig.\ref{scheme}). The cycle restarts with the next pump pulse. Each stroke is described in detail below.

\begin{figure}[htbp]
  \centering
  \includegraphics[width=0.7\textwidth]{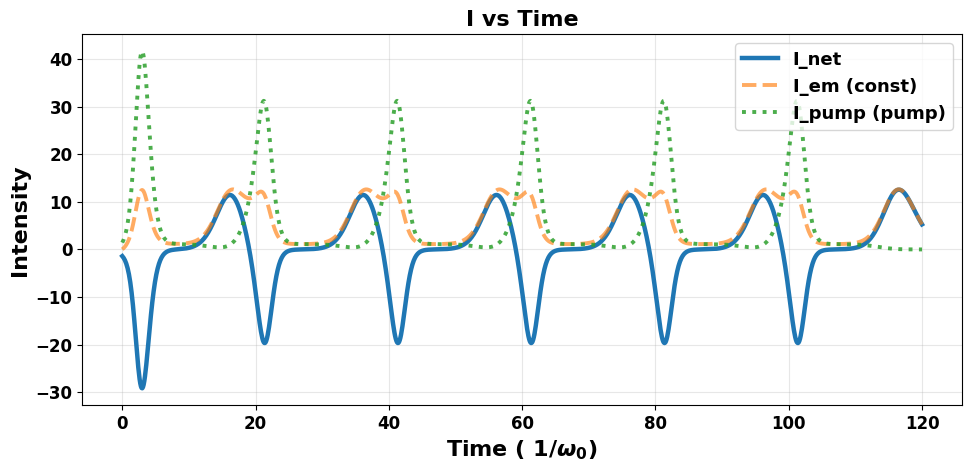}
  \caption{Superpulses across $k=5$ cycles. 
  Dotted (green) curve: pumping on; dashed (yellow) curve: natural decay with pumping off; 
  solid (blue) curve: resultant intensity during both superabsorption and superradiance strokes. 
  Input parameters (in units of $\omega_0$): 
  $N=80$, $T_c=0.5$, $\gamma_{\mathrm{down}}=0.01$, 
  $\gamma_{\mathrm{up}} = 3.5\,\gamma_{\mathrm{down}}$.}
  \label{fig:cycle_pulses}
\end{figure}

\begin{enumerate}

\item \textbf{Igniton: Cold Isochoric Thermalization} \\
    From $t=0$ to $t=t_{1}$, the system remains in contact with the bath at temperature $T_{c}>0$ ($\beta_{c}=1/T_{c}>0$), with fixed (representative) Hamiltonian $H(0) = H_{0} = \frac{\omega_{0}}{2}\,\sigma_{z},$ resulting in the thermal state $\rho_{1} = \frac{e^{-\beta_{c}H_{0}}}{Z_{0}}$.
    The vector radius and the initial angle on the Bloch sphere are determined by this state according to $n_z = \tanh\!\Bigl(\frac{\beta_c\,\omega_0}{2}\Bigr)$,
$r_{\mathrm{c}} = \lvert n_z\rvert,$
$\theta_c = \arccos(n_z).$ Note that in this ignition step no work is performed.

\
  \item \textbf{Unitary Superabsorbtion} \\
  
  \item From $t_1 $ to $t_{2}$ the system is collectively pumped. The Hamiltonian $H_0$ changes to Eq.(\ref{eq:hamiltonian}) with $\gamma_{\mathrm{down}}<\gamma_{\mathrm{up}}$ and 
    
    \[
      \rho_{1} \;\longrightarrow\; \rho_{2} = U\,\rho_{1}\,U^{\dagger}.
    \]

    At this stroke, the system undergoes a cyclic variation of its Hamiltonian $H_{0} \;\to\; H_{MF}(t) \;\to\; H_{0}$, in the time interval $\tau_{\rm abs}$ + $t_{d}^{\rm abs} \ll 1/\gamma_{\rm up}$. Here, $U$ is the evolution operator associated with $H_{MF}(t)$ given by Eq.(\ref{eq:hamiltonian}).
Although the Hamiltonian returns to its original form, the population becomes inverted and the system is characterized by an effective negative temperature. During this stroke, work $W_\text{pump}$ is done on the system through the pump.

  \item \textbf{{Unitary Superradiance}} \\
    
   In this stroke, the pumping is turned off and the system decays by superradiance. The Hamiltonian governing this stroke is given by Eq. (1),  with $\gamma_\text{\rm up}=0$, acting during the time interval $\tau_{\rm emis}$ + $t_{d}^{\rm emis} \ll 1/\gamma_{\rm down}$. Once again, the system undergoes a cyclic variation of its Hamiltonian, $H_{0} \to H_{\text{MF}}(t) \to H_{0}$. However, in this stroke, the state lowers its energy through the transformation $\rho_{2} \longrightarrow \rho_{3} = V \rho_{2} V^{\dagger}$, where $V$ is the evolution operator associated with $H_{\text{MF}}(t)$.  In this stroke work $W_\text{emis}$ is extracted from the system. The cycle is then restarted by turning the pump on. Note that after the ignition stroke, complete thermalization is not necessary; in this case, the superengine operates out of equilibrium and stabilizes from the second cycle onward (see Figs. ~\ref{fig:cycle_pulses} and ~\ref{fig:eta_vs_k}).

  \end{enumerate}

Two remarks are in order. First, external work is performed collectively during the superabsorption process by pumping the system. This pumping inverts the population and subsequently gives rise to superradiance. Second, work is extracted when the pumping is turned off and the system undergoes superradiant emission. Unlike conventional thermodynamic cycles (such as Otto or Carnot cycles), both superabsorption and superradiance occur while the transition frequencies of the $N$ two-level systems remain constant. However, the effective Hamiltonian $H(t)$ changes during superabsorption and superradiant pulses. Since these two strokes occur rapidly enough that the Liouvillian can be neglected see Appendix B, there is no heat exchange. Consequently, all work results exclusively from changes in the system Hamiltonian during these pulses. This represents the fundamental distinction between our proposed engine and conventional thermal engines.

\medskip{}

\subsection{Work and Power Scaling of the Superengine} In the unitary strokes, the entire change in internal energy results from work: specifically, the work $W_{\mathrm{pump}}$ done on the system by collective pumping  that induces superabsorption, and the work $W_{\mathrm{emis}}$ extracted from the system during superradiance. Thus, the extracted and the input work correspond to the area under the superradiant pulse $I(t)$ versus time and the superabsorption pulse, respectively. These values can be calculated analytically using Eq.~(\ref{eq:intensidade_superabsorcao}) of Appendix B, or obtained numerically from the exact model, and are shown to be proportional to $N$.

The average power output of the cycle is defined as $P = W_{\text{emis}} / \tau_{\text{cyc}}$, where $W_{\text{emis}}$ is the net work extracted and $\tau_{\text{cyc}}$ is the total cycle duration. This duration is approximately given by $\tau_{\text{cyc}} = t_{\text{therm}} + \tau_{\text{abs}} + \tau_{\text{emis}} + t_{d}^{\text{emis}} + t_{d}^{\text{abs}}$, where $t_{\text{therm}}$ is the thermalization time, $\tau_{\text{abs}}$ is the duration of the superabsorption stroke, $\tau_{\text{emis}}$ is the duration of the superradiance stroke, and $t_d$ is the delay time of the pulse. By switching the pump off (on) immediately after the superabsorption (superradiance) processe, the thermalization time can be ignored. Since each absorption and emission pulse has a characteristic width $\tau_{\text{abs}},\ \tau_{\text{emis}} \propto 1/N$, it follows that the total cycle time scales as $\tau_{\text{cyc}} \propto 1/N$. Combining this with the $N$-scaling of the work, we obtain:
\begin{equation}\label{eq:power_scaling}
P = \frac{W_{\text{net}}}{\tau_{\text{cyc}}} \propto \frac{N}{1/N} = N^2.
\end{equation}
Therefore, when the pulse durations are optimally tuned with $N$, the power of the superengine scales \emph{quadratically} with the number of emitters.

\medskip{}

\subsection{Efficiency} The figure of merit $\eta^{(k)}$ of this superengine for each cycle k  can be defined by the ratio of the energy extracted from the system to the energy invested, or, equivalently: 

\begin{equation}
\eta^{(k)}\;=\;\frac{\text{useful output}}{\text{supplied energy}}\;=\;\frac{W_{{\rm em}}^{(k)}}{W_{{\rm pump}}^{(k)}},
\end{equation}
where both $W_{\rm extr}$ and $W_{\rm pump}$ scale with $N$. 

To enhance the realism of the simulation, we implemented a smooth switching function for pump activation and deactivation, with the instantaneous on/off scenario represented by a staircase function in the limiting case. Furthermore, we kept the decay rate, $\gamma_{\text{down}}$, constant while scaling the pump rate according to $\gamma_{\text{up}} = x \gamma_{\text{down}}$.

\begin{figure}[htbp]
  \centering
  \includegraphics[width=0.8\textwidth]{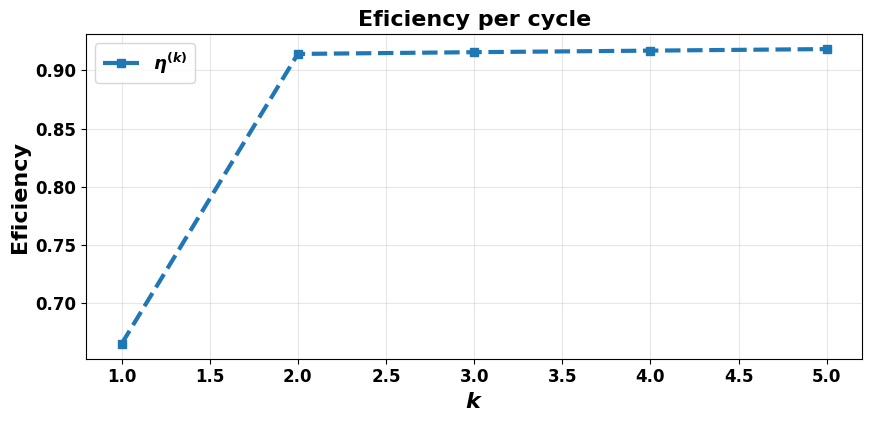}
  \caption{Efficiency $\eta$ versus number of cycles $k$. 
  The high efficiency is a hallmark of the superengine. 
  Input parameters (in units of $\omega_0$): 
  $N=80$, $T_c =0.5$, $\gamma_{\mathrm{down}}=0.01$, 
  $\gamma_{\mathrm{up}} = 3.5\,\gamma_{\mathrm{down}}$.} 
  \label{fig:eta_vs_k}
\end{figure}

The efficiency of a superabsorption-superemission cycle exhibits a non-trivial dependence on several parameters. For example, an excessively short pump switching time induces abrupt, non-adiabatic dynamics that generate significant entropy, drastically reducing operational efficiency. Conversely, a very long switching time, although more reversible, permits excessive energy loss through leakage via the natural decay channel $\gamma_{\text{down}}$ during the extended transition, which also lowers efficiency. Similarly, the pumping stroke duration, dissipation rates, pump strength $x$, and cold bath temperature—as detailed in the Appendix D- critically influence performance. Consequently, achieving peak operational efficiency requires tuning all parameters to an intermediate regime that carefully balances reversibility against entropy production.
Next, we perform an exact simulation to numerically calculate the efficiency for a cycle operated $k = 4$ times, as shown in Fig. \ref{fig:eta_vs_k}. This figure plots the efficiency $\eta$ against the number of cycles $k$ for $N = 80$. The rapid convergence of the efficiency demonstrates the stability of the high-performance regime. Using these same parameters, in Fig. \ref{fig:cycle_pulses}, we display the superabsorption and superradiance pulses as a function of the number of cycles $k=5$, using, in units of $\omega_0$: $N=80$, $T_c =0.5$, $\gamma_{\mathrm{down}}=0.01$, $\gamma_{\mathrm{up}}$= $3.5  \gamma_{{\rm down}}$.

As a final remark, we note that the parameters used in our simulations, correspond to a physical regime that can be implemented in experimental platforms operating under engineered reservoirs. 
In particular, the condition $\gamma_{\mathrm{up}} > \gamma_{\mathrm{down}}$ effectively describes a pumped reservoir with population inversion, as realized in optically or electrically driven quantum systems.  Moreover, rates as $\gamma_{\mathrm{up}}, \gamma_{\mathrm{down}} \ll \omega_{0}$ as we have used can be achieved in a variety of platforms, including collectively driven solid-state quantum emitters (e.g. quantum dots), superconducting qubits coupled to low-$Q$ microwave resonators, or plasmonic and excitonic systems with engineered broadband reservoirs \cite{liu2024super,lambert2016superradiance,tiranov2023collective,wei2021plasmon}. In these contexts, the chosen values of $\gamma_{\mathrm{up}}$ and $\gamma_{\mathrm{down}}$ represent experimentally accessible incoherent pumping and decay rates.

\section{Conclusion.}
In this work, we have introduced and analyzed a quantum engine cycle that exploits cooperative superabsorption and superradiance. Since the duration of each stroke scales as \(1/N\), adjusting the total cycle time accordingly enables the average power output to scale as 
$P \propto \frac{N}{1/N} = N^2$.
By mapping the full many-body dynamics onto an effective mean-field Hamiltonian with \(\sech^2\)-shaped drive pulses, we numerically demonstrate that the cycle can operate with efficiency arbitrarily close to unity. Furthermore, by iterating the cycle \(k\) times, we observe rapid convergence to a stable limit cycle, confirming the robustness of its performance under repeated operation. These results pave the way for scalable, ultra-efficient quantum heat engines that harness collective effects in realistic physical platforms.

\begin{appendix}

\section{A Single Two-Level System Driven by an Attenuator and an Amplifier}

Here we derive the Lindblad Master Equation and the corresponding steady-state population for a single two-level system (TLS) having a transition frequency $\omega$:
\[
H_S=\frac{\hbar\omega}{2}\,\sigma_z,
\qquad 
\sigma_\pm=\tfrac12(\sigma_x\pm i\sigma_y).
\]

Assume two independent reservoirs \cite{lorenzen2009quantum} where the attenuator ($m=1$) couples through the rotating term and the amplifier ($m=2$) couples through counter-rotating terms.
\;
The Hamiltonian is then
\begin{align}
H &= H_S + H_B + H_I(t), \nonumber\\
H_B &= \sum_{m=1}^{2}\sum_k \hbar\omega_{mk}\,b_{mk}^\dagger b_{mk}, \label{eq:HB}\\
H_I(t) &= \sum_k \Bigl[
g_{1k}\,(\sigma_- b_{1k}^\dagger + \sigma_+ b_{1k})
+ g_{2k}\bigl(e^{+i2\omega t}\sigma_- b_{2k} + e^{-i2\omega t}\sigma_+ b_{2k}^\dagger\bigr)
\Bigr].
\end{align}

\subsection*{Interaction Picture Hamiltonian}

We start from
\begin{align}
H &= H_S + H_B + H_I(t), \nonumber\\
H_B &= \sum_{m=1}^{2}\sum_k \hbar\omega_{mk}\,b_{mk}^\dagger b_{mk}, \label{eq:HB}\\
H_I(t) &= \sum_k \Bigl[
g_{1k}\,(\sigma_- b_{1k}^\dagger + \sigma_+ b_{1k})
+ g_{2k}\bigl(e^{+i2\omega t}\sigma_- b_{2k} + e^{-i2\omega t}\sigma_+ b_{2k}^\dagger\bigr)
\Bigr]. \nonumber
\end{align}

In the interaction picture, $H_I^{(I)}(t) = e^{\tfrac{i}{\hbar}H_0 t} H_I(t) e^{-\tfrac{i}{\hbar}H_0 t}$.
Using $\sigma_\pm(t)=e^{\pm i\omega t}\sigma_\pm$ and
$b_{mk}(t)=e^{-i\omega_{mk} t} b_{mk}$, $b_{mk}^\dagger(t)=e^{+i\omega_{mk} t} b_{mk}^\dagger$, we obtain
\begin{equation}
\begin{aligned}
H_I^{(I)}(t)
&= \sum_k g_{1k}\!\left[
\sigma_- b_{1k}^\dagger\, e^{\,i(\omega_{1k}-\omega)t}
+ \sigma_+ b_{1k}\, e^{-i(\omega_{1k}-\omega)t}
\right] \\[3pt]
&\quad + \sum_k g_{2k}\!\left[
\sigma_- b_{2k}\, e^{-i(\omega_{2k}-\omega)t}
+ \sigma_+ b_{2k}^\dagger\, e^{\,i(\omega_{2k}-\omega)t}
\right].
\end{aligned}
\label{eq:HI_int_picture}
\end{equation}

Introducing the detunings $\Delta_{mk} \equiv \omega_{mk}-\omega$, Eq.~\eqref{eq:HI_int_picture} reads
\begin{equation}
H_I^{(I)}(t)
= \sum_k g_{1k}\!\left[
\sigma_- b_{1k}^\dagger\, e^{\,i\Delta_{1k} t}
+ \sigma_+ b_{1k}\, e^{-i\Delta_{1k} t}
\right]
+ \sum_k g_{2k}\!\left[
\sigma_- b_{2k}\, e^{-i\Delta_{2k} t}
+ \sigma_+ b_{2k}^\dagger\, e^{\,i\Delta_{2k} t}
\right].
\label{eq:HI_detunings}
\end{equation}

Assume a weak coupling such that the baths remain thermal,
\(
\rho_{R}= \rho_{1}^{\text{th}}\!\otimes\rho_{2}^{\text{th}}.
\)
The second-order Nakajima–Zwanzig expansion yields
\begin{equation}
\frac{d\rho(t)}{dt}
=-\!\int_{0}^{\infty}\!\!d\tau\,
\Tr_{R}\!\bigl[H_I'(t),[H_I'(t-\tau),\rho(t)\otimes\rho_{R}]\bigr].
\label{eq:NZ}
\end{equation}

Identifying \( B_m(\tau)=\sum_k g_{mk} e^{-i(\omega_{mk}-\omega)\tau} b_{mk} \) (or \( b_{mk}^\dagger \)) we can define the correlation functions
\begin{align}
\int_{0}^{\infty}\!d\tau\,
\langle B_1^\dagger(\tau)B_1(0)\rangle &= \tfrac{\Gamma_1}{2}\,N_1, &
\int_{0}^{\infty}\!d\tau\,
\langle B_1(\tau)B_1^\dagger(0)\rangle &= \tfrac{\Gamma_1}{2}(N_1+1), \nonumber\\
\int_{0}^{\infty}\!d\tau\,
\langle B_2^\dagger(\tau)B_2(0)\rangle &= \tfrac{\Gamma_2}{2}(N_2+1), &
\int_{0}^{\infty}\!d\tau\,
\langle B_2(\tau)B_2^\dagger(0)\rangle &= \tfrac{\Gamma_2}{2}N_2,
\label{eq:correl}
\end{align}
where $\Gamma_m=2\pi\sum_k|g_{mk}|^2\delta(\omega-\omega_{mk})$ and 
$N_m(\omega)=\bigl[e^{\beta_m\hbar\omega}-1\bigr]^{-1}$.

Substituting \eqref{eq:correl} into \eqref{eq:NZ}, discarding the small Lamb shift,
and returning to the Schrödinger picture gives
\begin{equation}
\begin{aligned}
\dot{\rho} &=
-\frac{i\omega}{2}\,[\sigma_z,\rho]\;\\
&\quad+\gamma_{\rm down}\!\left(2\sigma_-\,\rho\,\sigma_+ - \{\sigma_+\sigma_-,\rho\}\right)\\[4pt]
&\quad+\gamma_{\rm up}\!\left(2\sigma_+\,\rho\,\sigma_- - \{\sigma_-\sigma_+,\rho\}\right),
\end{aligned}
\label{eq:master}
\end{equation}
with effective rates
\[
\gamma_{\rm down}= \Gamma_1(N_1+1)+\Gamma_2\,N_2,
\qquad
\gamma_{\rm up}= \Gamma_1 N_1+\Gamma_2(N_2+1).
\]

\subsection*{Rate equations and steady‐state populations}

Defining the excited-state population  \(
n_e(t)=\langle e|\rho(t)|e\rangle
      =\Tr\!\bigl(\sigma_+\sigma_-\,\rho(t)\bigr);
\)
the ground-state population 
\(n_g(t)=1-n_e(t)\),
and using the master equation~\eqref{eq:master} we obtain
\[
\frac{d n_e}{dt}
      =\Tr\!\bigl(\sigma_+\sigma_-\,\dot\rho\bigr)
      =\gamma_{\rm down}\,n_g-\gamma_{\rm down}\,n_e
      =-\bigl(\gamma_{\rm up}+\gamma_{\rm down}\bigr)n_e+\gamma_{\rm up},
\]
hence the assintotic solution reads
\[
n_e^{(\infty)}=\frac{\gamma_{\rm up}}{\gamma_{\rm up}+\gamma_{\rm down}},
\]
i.e.,\ whenever the \emph{effective gain} supplied by the amplifier exceeds the \emph{effective loss} induced by the attenuator, the two-level system supports population inversion. In the following, we extend this treatment to a sample of $N$ two-level emitters to obtain an effective mean fiel dynamics.

\section{Derivation of the Hamiltonian describing super radiance and superabsortion}

To obtain the Master Equation for \( N \) Spin-1/2 particles, our starting point is the many-body Hamiltonian 

\begin{equation}
\label{eq:H-full}
H = \omega_{0} S_{z}
    + \sum_{k}\omega_{k} b_{k}^{\dagger} b_{k}
    + \sum_{k}\omega_{k} c_{k}^{\dagger} c_{k}
    + \sum_{k}\lambda_{k}\!\left(S_{+} b_{k} + S_{-} b_{k}^{\dagger}\right)
    + \sum_{k}\zeta_{k}\!\left(S_{-} c_{k}\,e^{2 i\omega_{0} t}
                              + S_{+} c_{k}^{\dagger} e^{-2 i\omega_{0} t}\right),
\end{equation}
where the first term describes the collective spin of $N$ two-level systems (with
raising and lowering operators $S_{\pm}$ and inversion $S_{z}$).
The $b_{k}$ ($c_{k}$) modes represent two independent bosonic reservoirs that
mediate ordinary emission (absorption). Introducing the detuning $\Delta_k = \omega_k - \omega_0$ and moving the interaction picture, the Hamiltonian can be rewritten as
\begin{equation}
H_I(t) = \sum_k \lambda_k \!\left[
S_+ b_k\, e^{-i\Delta_k t} + S_- b_k^\dagger\, e^{+i\Delta_k t}
\right]
+ \sum_k \zeta_k \!\left[
S_- c_k\, e^{-i\Delta_k t} + S_+ c_k^\dagger\, e^{+i\Delta_k t}
\right].
\label{eq:HI_detuning}
\end{equation}

Note that counter-rotating interactions describe the Glauber amplifier \cite{mollow1967quantum,glauber1986amplifiers,grimaudo2019dynamics} and represent the collective pumping with amplitude $\zeta_k$ and phase $2\omega_0$, consisting of  $k$ modes and responsible for inverting population.

Tracing over both baths and applying the Born–Markov approximation yields the exact master equation

\begin{equation}
\dot{\rho}_{N} = -i\bigl[\omega_{0} S_{z},\,\rho_{N}\bigr]
                 + \mathcal{L}_{\mathrm{eff}}\rho_{N},
\end{equation}
with
\begin{align}
\mathcal{L}_{\mathrm{eff}}\rho_{N}
  &= \mathcal{L}[S_{-}]\,\rho_{N} + \mathcal{L}[S_{+}]\,\rho_{N},
  \\
\mathcal{L}[S_{\pm}]\,\rho_{N}
  &= \frac{\gamma_{\mathrm{eff}}}{2}\,(\bar n_{\mathrm{eff}}+1)
     \left(2S_{-}\rho_{N}S_{+} - S_{+}S_{-}\rho_{N} - \rho_{N}S_{+}S_{-}\right)
     \nonumber\\
  &\quad
     +\frac{\gamma_{\mathrm{eff}}}{2}\,\bar n_{\mathrm{eff}}
     \left(2S_{+}\rho_{N}S_{-} - S_{-}S_{+}\rho_{N} - \rho_{N}S_{-}S_{+}\right),
\end{align}
where
\[
\gamma_{\mathrm{eff}}
      = \gamma_{\rm down} - \gamma_{\rm up},
\qquad
\bar n_{\mathrm{eff}}
      = \frac{\gamma_{\rm down} n_{1} + \gamma_{\rm up}(n_{2}+1)}{\gamma_{\mathrm{eff}}}.
\]

Note that $\gamma_ {eff}$ is responsible for both superabsorption if $\gamma_ {\rm down} > \gamma_{\rm up}$ and superradiance if $\gamma_ {\rm down} < \gamma_{\rm up}$.

\subsection*{Single-spin (mean-field) dynamics}

Defining the reduced single-spin state $\rho \;=\; \operatorname{Tr}_{2,\dots,N}\!\rho_{N},$
and using $\operatorname{Tr}_{2,\dots,N}[S_{z},\rho_{N}]
     = [\omega_{0}s_{z}^{(1)},\rho]$,
the equation of motion becomes
\begin{equation}
\dot{\rho} = -i\,[H_{\mathrm{MF}}(t),\rho] + \mathcal{L}\rho,
\end{equation}
with the mean-field Hamiltonian, disregarding the Lamb-shift energy, is given by Eq.(1) of the main text, that is
\begin{equation}
H_{\mathrm{MF}}(t)
  = \frac{\omega_{0}}{2}\sigma_{z}
    + \frac{N}{2}\!\left(\gamma_{\rm down}-\gamma_{\rm up}\right)
      \bigl(\langle\sigma_{x}\rangle \sigma_{y}
           -\langle\sigma_{y}\rangle \sigma_{x}\bigr),
\end{equation}
where $\langle\sigma_{x}\rangle$, $\langle\sigma_{y}\rangle$ are calculated
self-consistently from~$\rho$, and the usual mean-field closure

$\sum_{r=p+1}^{N}\!\operatorname{Tr}_{p+1,\dots,N}
 \sigma_{\pm}^{(r)}\rho_{N} \approx (N\!-\!p)\,\operatorname{Tr}_{p+1,\dots,N}
 \sigma_{\pm}\rho_{N}$
has been employed \cite{mizrahi1993pulsed}. In Bloch representation,
$\langle\sigma_x\rangle = r\sin\theta\cos\phi,\quad \langle\sigma_y\rangle = r\sin\theta\sin\phi,\quad \langle\sigma_z\rangle = r\cos\theta$, 
or explicitly:
\begin{align}
\langle \sigma_x(t) \rangle
&= r \,\sech\!\Bigl(\tfrac{t - t_d}{\tau}\Bigr)
  \cos\bigl(\phi_0 + \omega_0 t\bigr),
  \label{eq:sigmax} \\[6pt]
\langle \sigma_y(t) \rangle
&= r \,\sech\!\Bigl(\tfrac{t - t_d}{\tau}\Bigr)
  \sin\bigl(\phi_0 + \omega_0 t\bigr),
  \label{eq:sigmay} \\[6pt]
\langle \sigma_z(t) \rangle
&= r \,\tanh\!\Bigl(\tfrac{t - t_d}{\tau}\Bigr),
  \label{eq:sigmaz}
\end{align}
where $\phi_0$ is the initial phase, which we choose to be zero,  $\omega_0$ is the atomic transition frequency, $t_d = \tau \ln\left(\cot(\theta_0/2)\right)$ is the delay time, marking the peak of the pulse, $\tau=2/rN|\gamma_{\rm eff}|$ is the characteristic time scale for superradiance ($\gamma_{\rm down} > \gamma_{\rm up}$) or superabsorption ($\gamma_{\rm up} > \gamma_{\rm down}$). The scaling factor $r$ locates the eigenstates of $H_{MF}(t)$ on the Bloch sphere oriented by angles $(\theta_{0},\phi_{0})$ and accounts for possible mixed-state effects, typically $r = 1$ for pure states. This effective Hamiltonian describes both a superradiant and a superabsorbing pulse, whose intensity can be promptly calculated and scales with $N^2$:

\begin{equation}
\mathcal{I}(t) = -N\frac{d\varepsilon(t)}{dt}
= \left(\frac{N}{2}\right)^2 r \bigl(\gamma_{\rm eff}\bigr) \omega_{0} \, \sech^2\!\Bigl(\frac{t - t_{d}}{\tau}\Bigr).
\label{eq:intensidade_superabsorcao}
\end{equation}

This approach simplifies the collective superradiance problem into a single atom coupled to a self-consistent mean field. Although it is an effective model, it accurately captures the key transient features of superradiance before atomic relaxation occurs: the characteristic delay time, the peak intensity, and the shape of sech²(t)  pulse. Consequently, it provides a useful framework for investigating both superradiant and superabsortion bursts.

The single-spin dissipator-attenuator $\mathcal{L}\rho$ in Eq.(11) is given by
\begin{align}
\mathcal{L}\rho
  &= \frac{\gamma_{\mathrm{eff}}}{2}\,(\bar n_{\mathrm{eff}}+1)
     \left(2\sigma_{-}\rho\sigma_{+}
           -\sigma_{+}\sigma_{-}\rho
           -\rho\sigma_{+}\sigma_{-}\right)
     \nonumber\\
  &\quad
     +\frac{\gamma_{\mathrm{eff}}}{2}\,\bar n_{\mathrm{eff}}
     \left(2\sigma_{+}\rho\sigma_{-}
           -\sigma_{-}\sigma_{+}\rho
           -\rho\sigma_{-}\sigma_{+}\right).
\end{align}
where the $\sigma_{\pm}$ and $\sigma_{z}$ act on a single emitter. Importantly, note that the unitary term $H_{MF}$ carries the number $N$ of two-level emitters. This means, as stated in the main text, that either for very short times or for a sufficiently large $N$, the dissipation term can be ignored.

\section{Description of the numerical (exact) treatment}

The simulation of the quantum thermodynamic cycle as described in the main text is composed essentially of two strokes. Indeed, after the initial thermalization stroke, we can adjust the strokes time duration such that \emph{superabsorption} and a \emph{superradiance} alternate. The working medium consists of $N$ identical two-level systems (qubits) coupled collectively to a external reservoir. The dynamics are modeled within the Lindblad-Markov framework using the \texttt{QuTiP} library \cite{johansson2012qutip}.

The collective spin operators are defined as
\begin{equation}
J_x = \frac{1}{2} \sum_{n=1}^N \sigma_x^{(n)}, \quad
J_y = \frac{1}{2} \sum_{n=1}^N \sigma_y^{(n)}, \quad
J_z = \frac{1}{2} \sum_{n=1}^N \sigma_z^{(n)},
\end{equation}
where $\sigma_\alpha^{(n)}$ denotes the Pauli matrix $\sigma_\alpha$ acting on the $n$-th qubit and as the identity on all others. The collective raising and lowering operators are
\begin{equation}
J_+ = J_x + i J_y, \quad J_- = J_x - i J_y,
\end{equation}
and the system Hamiltonian is given by
\begin{equation}
H = \omega_0\, J_z,
\end{equation}
where $\omega_0$ is the energy splitting between the two levels.

\subsection*{Master Equation}

The time evolution of the system's density matrix $\rho(t)$ is governed by the standard Lindblad master equation
\begin{equation}
\frac{d\rho}{dt} = -i[H, \rho] + \sum_j \mathcal{D}[c_j]\,\rho,
\label{eq:lindblad}
\end{equation}
with the dissipator defined as
\begin{equation}
\mathcal{D}[c]\,\rho = c\,\rho\,c^\dagger - \frac{1}{2} \left( c^\dagger c \,\rho + \rho\, c^\dagger c \right).
\end{equation}

The collapse operators $c_j$ depend on the stroke of the cycle:
\begin{itemize}
    \item \textbf{Superabsorption stroke:}
    \begin{equation}
    c_{\mathrm{abs}} = \sqrt{\gamma_{\mathrm{abs}}}\, J_+,
    \end{equation}
    leading to
    \begin{equation}
    \frac{d\rho}{dt} = -i[H,\rho] + \gamma_{\mathrm{abs}}\, \mathcal{D}[J_+]\,\rho.
    \end{equation}
    \item \textbf{Superradiance stroke:}
    \begin{equation}
    c_{\mathrm{em}} = \sqrt{\gamma_{\mathrm{em}}}\, J_-,
    \end{equation}
    leading to
    \begin{equation}
    \frac{d\rho}{dt} = -i[H,\rho] + \gamma_{\mathrm{em}}\, \mathcal{D}[J_-]\,\rho.
    \end{equation}
\end{itemize}
If pure dephasing is included, an additional collapse operator
\begin{equation}
c_{\phi} = \sqrt{\gamma_{\phi}}\, J_z
\end{equation}
is added to Eq.~\eqref{eq:lindblad}.

\begin{figure}[htbp]
  \centering
  \includegraphics[width=0.8\linewidth]{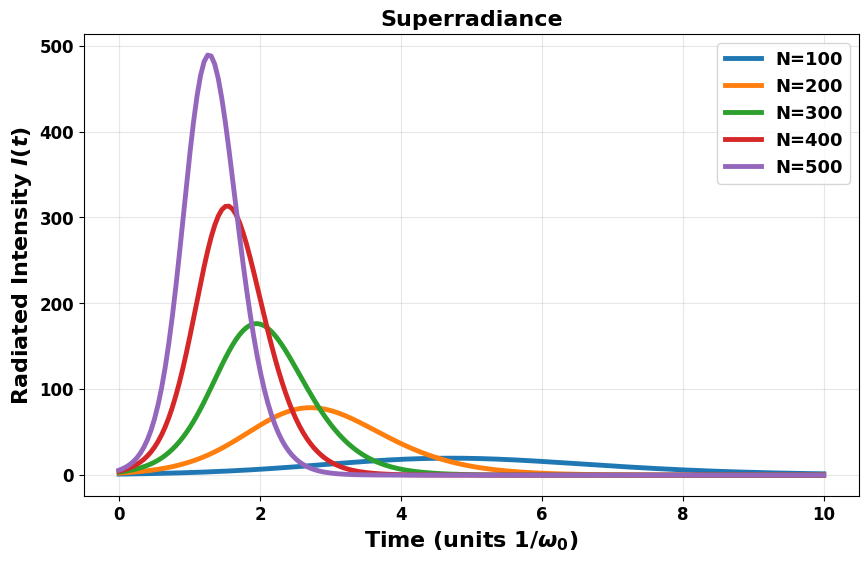}
  \caption{Superradiance curves for several $N$.}
  \label{fig:sr_curves}
\end{figure}

\begin{figure}[htbp]
  \centering
  \includegraphics[width=0.8\linewidth]{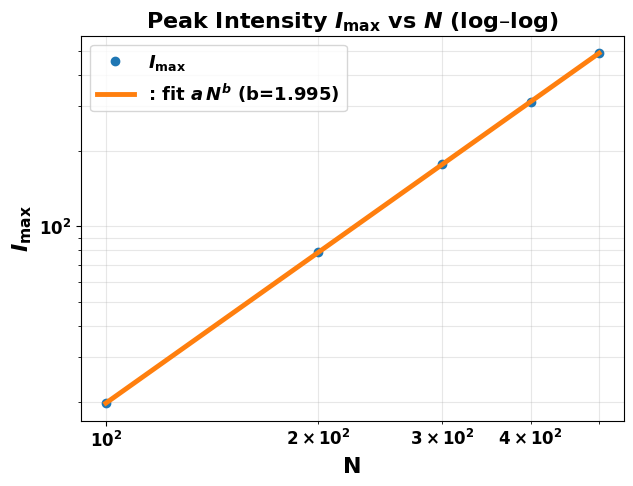}
  \caption{Fitting curve for maximum intensity versus $N$.}
  \label{fig:sr_fit}
\end{figure}

As an initial state, we used a collective thermal state since, due to the Hamiltonian we use, the dynamic is restricted to the symmetric subspace with total angular momentum $J = N/2$ \cite{gross1982superradiance}, in which case $\rho_\beta^{(\mathrm{sym})} \propto \sum_{m=-J}^J e^{-\beta \omega_0 m} \, |J,m\rangle\langle J,m|$. This reduces the Hilbert space from from $2^N$ to $N+1$, allowing simulations with much larger N. As observed from our numerical simulations, the peak intensity scales with $N^b$, with b approaching the value 2 as N increases,  as shown in Fig.~\ref{fig:sr_fit}.

\subsection*{Cycle Description}
As described in the main text, to avoid having to account for the work of preparing states, the simulation begins with the collective thermal state of $N$ identical qubits at a reference cold temperature $T_c = 1/\beta$. Next, the system evolves under $c_{\mathrm{abs}}$, collectively absorbing energy from a cold bath and increasing its excitation. The cycle completes with superradiance, when the system evolves under $c_{\mathrm{em}}$, collectively releasing energy to the cold bath.

In each stroke, the intensity-like observables $\langle J_-J_+\rangle$ (absorption) and $\langle J_+J_-\rangle$ (emission) are recorded, allowing the computation of the absorbed and emitted work via numerical integration.

\subsection*{Efficiency}
To enhance the realism of the simulation, we implemented a smooth switching function for the pump activation and deactivation. The function $
S(t) = \frac{1}{2} \left[ 1 + \tanh\left( \frac{t - t_0}{\tau_{\text{on/off}}} \right) \right]$
interpolates between the instant the pump is turned on or off, where \(t_0\) denotes the center of a specific transition (on/off) and \(\tau_{\text{on/off}}\) controls the times cale (abruptness) of the transition. The sudden on/off scenario is recovered in the limiting case of \(\tau_{\text{on/off}} \to 0\), where \(S(t)\) approaches a staircase function. Note that the emission channel remains always active, i.e., \(\gamma_{\text{down}}\) is constant. In our code, this function modulates the collective pumping rate as $
\gamma_{\text{pump}}(t) = x \, S(t) \, \gamma_{\text{down}}, \quad x > 0$,
which acts as a smooth temporal envelope during each cycle.

From the numerically evaluated expectation values
($\langle\cdot\rangle_t \equiv \Tr[\cdot\,\rho(t)]$), we define the instantaneous intensities
\begin{equation}
I_{\rm em}(t) \;=\; \gamma_{\rm em}\,\big\langle J_+J_- \big\rangle_t,
\qquad
I_{\rm pump}(t) \;=\; -\,\gamma_{\rm pump}(t)\,\big\langle J_-J_+ \big\rangle_t,
\qquad
I_{\rm net}(t) \;=\; I_{\rm em}(t)+I_{\rm pump}(t).
\end{equation}
We adopt the convention $I_{\rm em}(t)\!>\!0$ as \emph{output} (useful work) and
$I_{\rm pump}(t)\!<\!0$ as \emph{input} (pumping cost). For cycle $k$,
the corresponding works are as follows

\begin{align}
W_{\rm pump}^{(k)} &= \int_{\mathcal{I}^{(k)}_{\rm abs}} \!\! \bigl(-I_{\rm pump}(t)\bigr)\,dt
\;=\; \int_{\mathcal{I}^{(k)}_{\rm pump}} \!\! \gamma_{\rm pump}(t)\,\langle J_-J_+\rangle_t \, dt,
\\
W_{\rm em}^{(k)}  &= \int_{\mathcal{I}^{(k)}_{\rm em}} \!\! I_{\rm em}(t)\,dt
\;=\; \int_{\mathcal{I}^{(k)}_{\rm em}} \!\! \gamma_{\rm em}\,\langle J_+J_-\rangle_t \, dt.
\end{align}
Since emission cannot be switched off during the absorption half-stroke, there is an
\emph{inevitable leakage},
\begin{equation}
W_{\rm leak}^{(k)} \;=\; \int_{\mathcal{I}^{(k)}_{\rm abs}} \!\! I_{\rm em}(t)\,dt
\;=\; \int_{\mathcal{I}^{(k)}_{\rm abs}} \!\! \gamma_{\rm em}\,\langle J_+J_-\rangle_t \, dt,
\end{equation}
which represents emission (loss) while the system is being pumped. The figure of merit that best describes this cycle and is experimentally accessible in each cycle $k$ is the following

\begin{equation}
\eta^{(k)}\;=\;\frac{\text{useful output}}{\text{supplied energy}}\;=\;\frac{W_{{\rm em}}^{(k)}}{W_{{\rm pump}}^{(k)}}.
\end{equation}

Note that this is the efficiency that can be \emph{actually observed and measured} in an experiment, since it accounts for the full energetic cost of the real pumping process.

\section{Entropy production impacting on Efficiency}

As we have said, pump activation and deactivation, as well as its increasing, are described by smooth $S(t)$, which is
$\tanh$-shaped windows with a characteristic switching time $\tau_{\rm on/off}$ (not to be confused with the mean width of the superradiante/absorption pulses). 
This parameter controls how rapidly the collective absorption channel 
is turned on or off during each half-cycle.
When $\tau_{\rm on/off}$ is small, the pump is switched almost instantaneously. 
This sudden change injects energy into the system in a highly 
non-adiabatic manner, which leads to larger entropy production thus reducing, efficiency. 
When $\tau_{\rm on/off}$ is large, the pump amplitude increases and decreases 
more gradually, and the absorption stroke approaches 
quasi-reversible dynamics, thus improving efficiency. 
However, because the emission channel ($\gamma_{\rm down}$) is always active, 
a longer switching time also implies that more energy is lost 
by leakage during the pumping stage and may lower the efficiency as well. Sumarizing, pump switching time $\tau_{\rm on/off}$ must be tuned to an 
intermediate regime: smooth enough to suppress irreversibility, 
but not so long that leakage becomes the main source of inefficiency. 

Another parameter that affects efficiency is the time duration $\tau_{\rm stroke}$ of each half-stroke of the cycle: one 
absorption stroke with the pump active and one emission stroke with the pump inactive. When this time is short, the cycle is fast but highly irreversible. 
The system cannot reach quasi-equilibrium, and entropy production grows, thus decreasing $\eta_{\rm pump}$. On the other hanc, when this time is long the absorption becomes nearly reversible, 
suppressing entropy production and increasing efficiency. Nevertheless, since the emission channel is always active, the leakage $Q_{\rm leak}$ grows proportionally  to this time, 
which eventually reduces $\eta_{\rm pump}$. Longer strokes also reduce the output power. 

The dissipation  $\gamma_{\rm down}$ and pumping $\gamma_{\rm up}$ rates also affects the efficiency. In fact, the emission rate $\gamma_{\rm down}$ is always active, setting the baseline leakage. The peak absorption rate is $\gamma_{\rm up}^{\rm peak} = x \gamma_{\rm down}$. In general, larger $\gamma_{\rm down}$ increases losses throug leaking and therefore lowers  
efficiency, while larger $\gamma_{\rm up}$ enables stronger excitation, which increases the work extracted $W_{\rm em}$, but if too strong, it induces non-adiabaticity and entropy production, reducing efficiency. 

Regarding the parameter $x$, it is important to bear in mind that while $x$ can be chosen freely at the level of the numerical simulation, its physical validity range is restricted by the approximations underlying the Lindblad master equation employed here. First, the master equation is derived under the \emph{Born–Markov approximation}, which requires that the system--reservoir coupling is weak compared to the system's characteristic frequency scale $\omega_{0}$. In practice, this translates into the condition $\gamma_{\text{\rm down}}, \, \gamma_{\text{\rm up}}^{\text{peak}} \ll \omega_{0}$.
If this inequality is violated, the environment correlation time becomes comparable to the system timescales, and memory effects (non-Markovianity) can no longer be neglected. Second, the \emph{Rotating Wave Approximation} (RWA), used in deriving the collective jump operators $J_{p}$ and $J_{m}$, also presupposes that the dissipative rates are small compared to $\omega_{0}$. Otherwise, counter-rotating terms that are neglected in the RWA may contribute significantly to the dynamics. Thus, the theoretical framework is only consistent provided $x \, \gamma_{\text{\rm down}} \ll \omega_{0}$.
For typical values such as those used in our code, $\gamma_{\text{em}} = 0.01 \, \omega_{0}$, this implies a safe regime of validity up to about $x \lesssim 10$, corresponding to $\gamma_{\text{\rm up}}^{\text{peak}} \lesssim 0.1 \, \omega_{0}$. Pushing $x$ significantly beyond this bound may still be numerically feasible, but the results cannot be interpreted within the standard RWA/Born–Markov formalism, as additional physical effects (non-secular terms, non-Markovian memory, or strong-coupling corrections) would need to be taken into account.

\section{Numerical implementation (discretization)}
In our code, the evolution is calculated with \texttt{mesolve} (QuTiP) using the collapse
operators $\sqrt{\gamma_{\rm em}}\,J_-$ (constant emission) and
$\sqrt{\gamma_{\rm abs}(t)}\,J_+$ (pumping modulated by $w_{\rm abs}(t)$).
At each time grid point $t_n\in{\tt tlist}$ we read
$\langle J_-J_+\rangle_{t_n}$ and $\langle J_+J_-\rangle_{t_n}$ to form
$I_{\rm abs}(t_n)$, $I_{\rm em}(t_n)$ and $I_{\rm net}(t_n)$.
The defining integrals for $W_{\rm abs}^{(k)}$, $W_{\rm em}^{(k)}$, and
$W_{\rm leak}^{(k)}$ are approximated by the trapezoidal rule
(\texttt{numpy.trapz}) \emph{restricted} to the time windows
$\mathcal{I}^{(k)}_{\rm abs}$ and $\mathcal{I}^{(k)}_{\rm em}$.

\subsection*{Algorithm Flow}
The algorithm implemented in the code can be summarized as follows:

\begin{enumerate}
    \item Define the system parameters ($N$, $\omega_0$, $\gamma_{\mathrm{up}}$, $\gamma_{\mathrm{down}}$, $\beta$, $t_{\max}$, $n_{\mathrm{points}}$).
    \item Construct collective spin operators $J_z$, $J_+$, $J_-$.
    \item Build the Hamiltonian $H = \omega_0 J_z$.
    \item Prepare the initial state: collective of $N$ qubit thermal states at temperature $T_c$.
    \item For each cycle $k = 1, \dots, K$:
    \begin{enumerate}
        \item Evolve the state under the superabsorption master equation, store the final state.
        \item Compute $I_{\mathrm{abs}}(t)$ and integrate to obtain $W_{\mathrm{pump}}$.
        \item Evolve the state under the superemission master equation, store the final state.
        \item Compute $I_{\mathrm{em}}(t)$ and integrate to obtain $W_{\mathrm{em}}$.
        \item Evaluate the efficiency $\eta^{(k)} = \frac{W_{\mathrm{em}}^{(k)}}{W_{\mathrm{pump}}^{(k)}}$.
        \item Update the state for the next cycle.
        \item Plot $\eta^{(k)}$ vs.\ $k$.
    \end{enumerate}
\end{enumerate}

\subsection*{Acknowledgements}
We acknowledge financial support from the Brazilian agencies: Coordenação de Aperfeiçoamento de Pessoal de Nível Superior (CAPES), financial code 001 and CNPq - Conselho Nacional de Desenvolvimento e Pesquisa, Grant 304028/2023-1. NGA and MHYM thank FAPESP Grant 2024/21707-0 and 2024/13689-1.

\end{appendix}

\bibliographystyle{apsrev4-2}
\bibliography{references}
\end{document}